\documentclass[conference]{IEEEtran}
\IEEEoverridecommandlockouts
\usepackage{cite}
\usepackage{amsmath,amssymb,amsfonts}
\usepackage{algorithmic}
\usepackage{graphicx}
\graphicspath{{figs/}}
\usepackage{cite}
\usepackage{textcomp}
\usepackage{xcolor}
\def\BibTeX{{\rm B\kern-.05em{\sc i\kern-.025em b}\kern-.08em
    T\kern-.1667em\lower.7ex\hbox{E}\kern-.125emX}}
\begin{document}

\title{An Improved Automatic Modulation Classification Scheme Based on Adaptive Fusion Network
\thanks{This work is supported by the National 111 Center.}
}

\author{\IEEEauthorblockN{1\textsuperscript{st} Hao Shi}
\IEEEauthorblockA{\textit{School of Microelectronics} \\
\textit{Xidian University}\\
Xi'an, China \\
852471751@qq.com}
\and
\IEEEauthorblockN{2\textsuperscript{nd} Qi Peng \thanks{\IEEEauthorrefmark{1} Qi Peng is corresponding author.}\IEEEauthorrefmark{1} }
\IEEEauthorblockA{\textit{School of Microelectronics} \\
\textit{Xidian University}\\
Xi'an, China \\
qipeng08@gmail.com}
\and
\IEEEauthorblockN{3\textsuperscript{rd} Yiqi Zhuang}
\IEEEauthorblockA{\textit{School of Microelectronics} \\
\textit{Xidian University}\\
Xi'an, China \\
yqzhuang@xidian.edu.cn}
}

\maketitle

\begin{abstract}
Due to the over-fitting problem caused by imbalance samples, there is still room to improve the performance of data-driven automatic modulation classification (AMC) in noisy scenarios. By fully considering the signal characteristics, an AMC scheme based on adaptive fusion network (AFNet) is proposed in this work. The AFNet can extract and aggregate multi-scale spatial features of in-phase and quadrature (I/Q) signals intelligently, thus improving the feature representation capability. Moreover, a novel confidence weighted loss function is proposed to address the imbalance issue and it is implemented by a two-stage learning scheme. 
Through the two-stage learning,  AFNet can focus on high-confidence samples with more valid information and extract effective representations, so as to improve the overall classification performance. In the simulations, the proposed scheme reaches an average accuracy of 62.66\% on a wide range of SNRs, which outperforms other AMC models. The effects of the loss function on classification accuracy are further studied.

\end{abstract}

\begin{IEEEkeywords}
Automatic modulation classification, deep learning, convolutional neural networks, loss function.
\end{IEEEkeywords}

\section{Introduction}
The evolving fifth generation (5G) communication systems are turning our society into a densly connected one, placing higher demands on communication speed and quality. Automatic Modulation Identification (AMC), as one of the key technologies for non-collaborative wireless communication systems, aims to identify the modulation scheme of the received signal with unknown channel interference \cite{b2}. Traditionally, AMC algorithms are divided into two categories: likelihood-based (LB) and feature-based (FB) \cite{b3}. Whereas the LB classifiers are theoretically optimal, they involve unacceptable computational costs when estimating unknown parameters. The FB methods have lower complexity, but the performance depends greatly on expertise and the quality of hand-crafted features.

Compared to traditional feature engineering, deep learning (DL) algorithms attempt to learn the deep feature hierarchies and directly induce the complex function mapping input to output from original data. In recent years, many DL-based neural networks have been designed in order to achieve high-performance AMC \cite{b5, b6,b7, b8, b9, b10}. O'Shea et al. proposed a compact CNN structure to identify eleven modulation schemes using baseband in-phase/quadrature (I/Q) samples \cite{b5}. Rajendran et al. trained a dual-layer long-short term memory (LSTM) using the amplitude/phase (A/P) representation of raw data to extract the long-term dependencies and improve the classification accuracy \cite{b7}. Some other studies combined spectrum analysis and neural networks to enrich the features by providing different modal information \cite{b8}. 

Although the above schemes vary in terms of data format and network design, they all follow the conventional DL training strategy that randomly sampling in each batch and the importance difference of samples are ignored. However, some samples contain strong noise and some contain invalid silent periods \cite{b5}. These abnormal samples are very difficult to classify and have a high probability of being outliers in the sample space. The neural network is easily to memorize the invalid knowledge if the training dataset includes a certain proportion of these samples, thus falling into local optimum and leading to degradation of classifier performance\cite{b14}. Furthermore, the network architectures are mostly migrated from the generic models and lack special design for extracting modulation information.
\begin{figure*}[htbp]
\centerline{\includegraphics[scale=0.75]{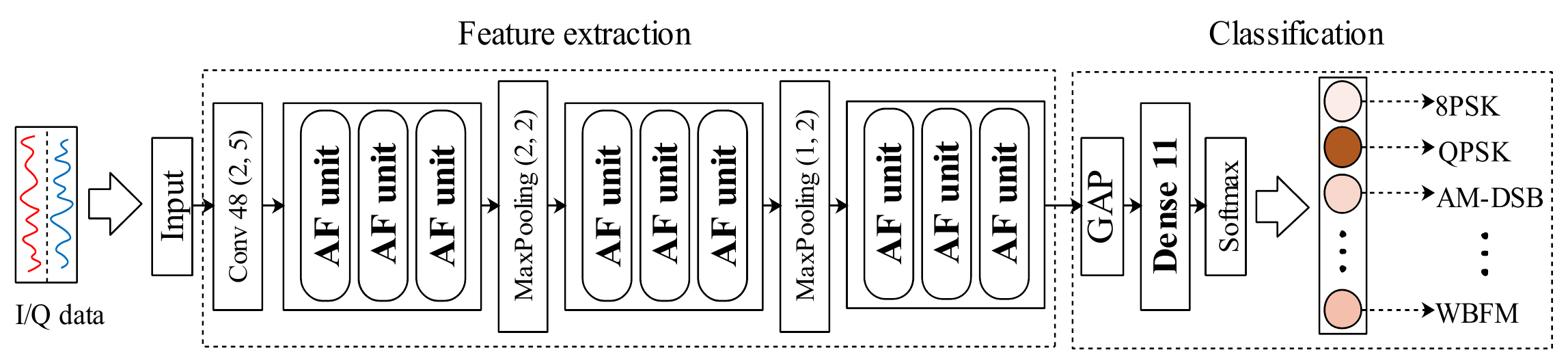}} 
\caption{The architecture of proposed AFNet.} 
\label{fig1} 
\end{figure*}

Based on the above considerations, this paper puts forward an improved scheme from two parts: classifier and loss function. First, the modulation information is usually contained in several adjacent symbols. Therefore, multi-scale features and dynamic receptive fields which can cover the key symbols are very important for the classifier. Thus, we propose a novel neural network with specially designed multi-scale parallel convolution and adaptive feature fusion to enrich the high-level features, namely adaptive fusion network (AFNet). Second, in order to alleviate the interference of the hard samples, per-instance weights are applied to the cross-entropy loss function based on the confidence level of each instance to guide the attention allocation for model training. The simulation results on RadioML dataset show that the combined AMC scheme can achieve superior overall accuracy from -20 dB to 18 dB than other DL-based models. The extended experiments evaluate that the proposed weighted loss is also effective for other models.

The rest of this paper is organized as follows. Section II presents the signal model. Section III introduces the proposed AMC scheme. Simulation results are presented and discussed in Section IV. Finally, Section V concludes the paper.



\section{Signal Model}


AMC can generally be regarded as the M-classification problem of complex base-band time series, where M represents the number of candidate modulation schemes. A general expression for the received signal is

\begin{equation}
r(t)=s(t)+n(t),
\label{1}
\end{equation}
where $s(t)$ represents noise-free baseband complex envelope of the received signal and $n(t)$ is the additive white Gaussian noise (AWGN) with a mean of zero and variance $\sigma_n^2$. 
The received signal $r(t)$ is sampled $N$ times to generate a discrete sequence $r[n], n=1,2,...,N$, which can be expressed as

\begin{equation}
r[n]=I[n]+jQ[n]=Re(r[n])+jIm(r[n]).
\label{2}
\end{equation}
The real and imaginary parts of $r[n]$ can be extracted and packed into a two-dimensional (2-D) matrix with shape $(2,N)$ and the \textit{k-th} collected sequence $r_k$ can then be written as follows:

\begin{equation}
{r}_k=\left( \begin{array}{c}
	Re(r_k[1],...,r_k[N])\\
	Im(r_k[1],...,r_k[N])\\
\end{array} \right) \text{.}
\label{3}
\end{equation}

For a given \textit{M}-class modulation scheme set $\mathbf{\Theta }=\{\theta _i\}_{i=1}^{M}$, according to the maximum \textit{a posteriori} probability criterion, the decision can be made as $
\theta _i=\mathop {\mathrm{arg}\max}_{\theta _i\in \Theta}P(\theta _i|{r}),
$, where  $
P(\theta _i|\boldsymbol{r})
$ is the a posteriori probability of $
\theta _i
$  under the condition of received sequence $r$.


\section{The Proposed AMC Scheme}


\begin{figure}[htbp]
\centerline{\includegraphics[scale=0.5]{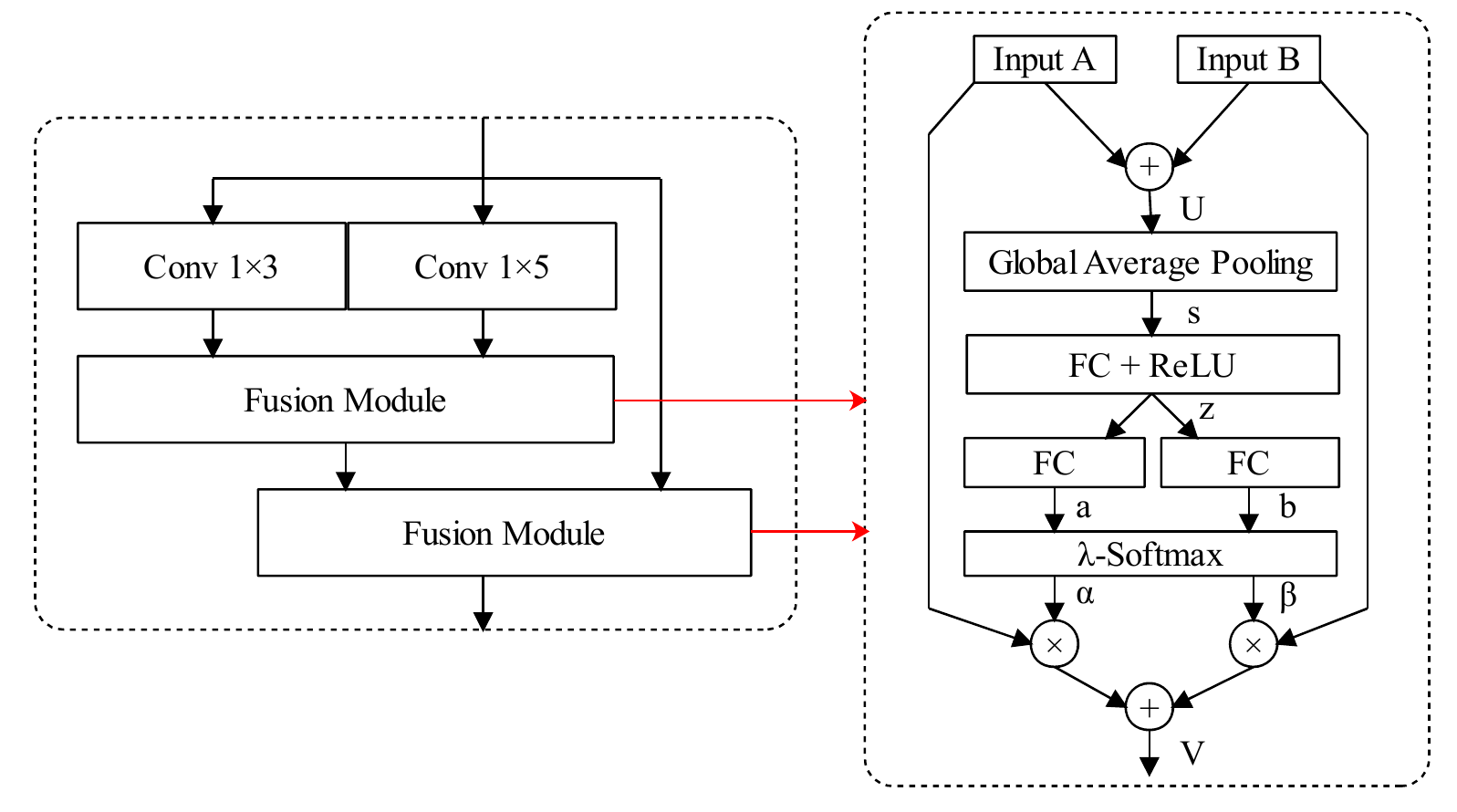}} 
\caption{The structure of AF unit.} 
\label{fig2} 
\end{figure}

\subsection{Implementation of AFNet}

Fig.~\ref{fig1} demonstrates the architecture of the proposed AFNet, which is composed of feature extraction stage and classification stage. The first layer is a conventional convolution layer to learn the general high-resolution features and the correlation between I and Q, which utilized 48 filters with size of $(2, 5)$. Then, this is followed by nine AF units, in which two max pooling layers are inserted to reduce the spatial dimensionality, as well as the computational complexity. The function of AF units is to extract and aggregate multi-scale features for next classification stage. In many DL related works, the fusion of multi-scale and multi-level features is one method for improving performance. There are two main fusion methods of feature maps: concatenation and element-wise summation. However, such linear superposition methods lack flexibility in the selection of features. Therefore, we designed an AF unit based on the attention mechanism to adaptively learn the weight vectors for feature maps and provide nonlinear fusion. 

The structure of AF unit is shown in Fig.~\ref{fig2}. In each unit, dual-branch parallel convolution is firstly set to capture small-scale and large-scale spatial features with kernel sizes of $(1, 3)$ and $(1, 5)$, respectively.  The number of filters is consistent with the channels of the input, in order to maintain the dimensionality after convolution. In actual implementation, we adopt the group convolution \cite{b11} and dilated convolution \cite{b12} to reduce the parameters. The right part of Fig.~\ref{fig2} is the schematic diagram of the fusion module based on attention mechanism. For given feature maps $A, B \in \mathbb{R}^{H\times W\times C}$, they are processed by sequential summation-global average pooling (GAP)-fully connections (FC) and converted into
two C-dimension vectors $a$ and $b$, and a 
modified channel-wise Softmax function with a factor $\lambda$ (referred as $\lambda$-Softmax) is applied to acquire the attention vectors $\alpha$ and $\beta$ as follows:

\begin{equation}
\alpha _c=\frac{\lambda e^{a_c}}{e^{a_c}+e^{b_c}},\beta _c=\frac{\lambda e^{b_c}}{e^{a_c}+e^{b_c}},
\label{9}
\end{equation}
where $c=0,1,...,C$ represents the channel index. In the first fusion module, the factor $\lambda$ is set to 1 and the function become the vanilla Softmax. While it is set to 2 in the second module which can be regarded as an optimized skip connection in ResNet. The main purpose of setting to 2 is to prevent vanishing gradient problems during the back propagation \cite{b13} and accelerate the training process. Finally, the output feature map $V={[V_1, V_2,...,V_C]}, V_c \in \mathbb{R}^{H\times W} $ is obtained  by the fusion of feature maps using the following formula:

\begin{equation}
V_c=\alpha _c\cdot A_c+\beta _c\cdot B_c,\alpha _c+\beta _c=\lambda.
\label{10}
\end{equation}
For each fusion module, only the FC layers involves trainable parameters, and the total number $N_{train}$ is

\begin{equation}
N_{train}=C\times d\times 3=3C^2/r.
\label{11}
\end{equation}
where $r$ is the compression factor of the FC layer and its default value is 16. Therefore, the amount of parameters generated by the attention module is only related to the number of channels, which is determined by the number of filters.

Through the nonlinear fusion of features maps from different branches, the AF unit can obtain dynamic receptive fields to adapt to different inputs. Similar to traditional convolution layers, the AF units can be flexibly embedded into any CNN structure. By stacking the units, the multi-scale features obtained from different layers are scaled and propagated through the entire network by the modified skip connections, thereby enriching the classification information. Considering the balance between accuracy and efficiency, the benchmark model utilizes nine AF units. 

In regard to the classifier, the parameters could be greatly cut down by GAP to transform the 2-Dimensional (2-D) feature map into a vector instead of a flattened layer followed by several FC layers. The vector is fully connected to the last layer with 11 neurons, which corresponds to the number of classes. Finally, a Softmax layer is applied to calculate the \textit{a posteriori} probability vector. 

Once the model is built, it can be trained on a dataset by using a loss function and optimization algorithm based on stochastic gradient descent (SGD). The training flow related to the proposed confidence weighted loss is described in detail in the following subsection.

\subsection{Pipeline of The Proposed Two-stage Learning}\label{AA}

\begin{figure}[t]
\centerline{\includegraphics[scale=0.4]{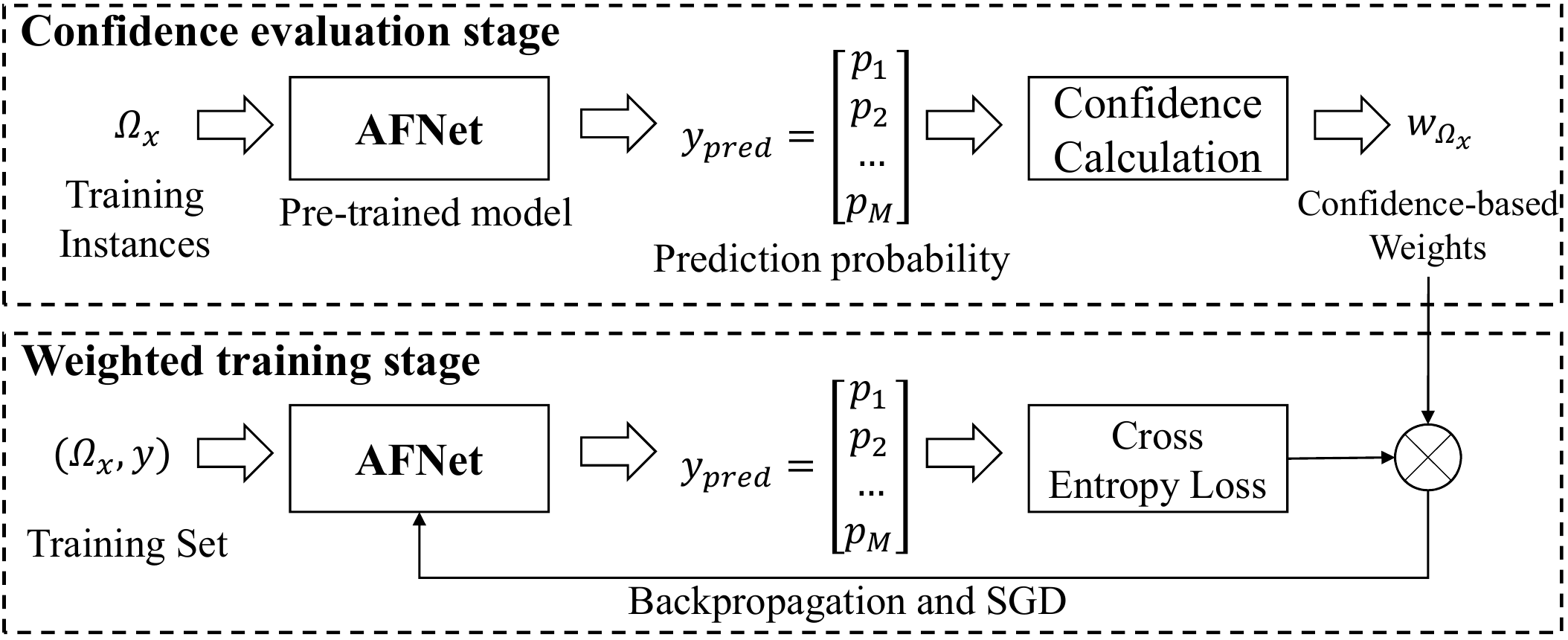}} 
\caption{The pipeline of proposed two-stage learning scheme.} 
\label{fig3} 
\end{figure}

In conventional DL-based AMC methods, the deep models are trained with general single-stage strategy utilizing the categorical cross entropy loss function 

\begin{equation}
L_{CE}=-\sum_{i=1}^{M}y_ilog(p_i)\text{,}
\label{4}
\end{equation}
where $M$ is the number of categories, $y$ represents the one-hot label and $p$ is the prediction of classifier, respectively. In the iterative training, the model is executed and the trainable parameters are continuously updated epoch by epoch until the loss converges to a constant. In one epoch, the training samples are shuffled and fed into the model in batches to calculate the batch loss and update the parameters. 

However, if there are a large number of low-quality samples in the dataset, the losses of them will account for the majority of the total loss. And most of these samples are difficult to classify, so the optimization direction of the model is not what we want. In order to solve such problems, the idea of curriculum learning is proposed driven by the human learning pattern, that is, the model should pay more attention to easy samples and gradually generalize to hard samples \cite{b14}. Inspired by this, we propose an instance confidence weighted loss (CW loss) function to suppress the hard samples and make the descending direction of gradient dominated by high-confidence samples. Moreover, the weights are given by the model itself rather than artificial design to ensure the reliability. The full pipeline of the proposed learning strategy consisting of two stages is shown in Fig.~\ref{fig3}.

The first stage namely confidence evaluation stage aims at computing the weights for training examples according to the predicted posterior probability distribution of the model. Firstly the model is pre-trained with training set $\Omega_x$ and label $y$ using conventional cross entropy loss. Then, the training instances are fed into the model again to obtain the prediction vector $y_{pred}$ for each instance. The elements $p_i, i=0, 1..., M$ in $y_{pred}$ is the probability of the \textit{i-th} class and $\sum_{i=1}^{M}p_i=1$. The final prediction result is determined by the largest item in the vector. However, the results given by the model may be ambiguous. For example, for a binary classification problem, the probability distributions of $[0.9, 0.1]$ and $[0.6, 0.4]$ will lead to a same prediction class, but the former has a higher confidence level. Therefore, a metric is needed to measure the uncertainty of prediction results. A common metric to identify uncertain examples is the entropy of their predicted class distribution \cite{b15}. For $p = [p_1, p_2,..., p_M]$, the entropy is defined as

\begin{equation}
H(p)=-\sum_{i=1}^{M}p_ilog(p_i)\text{.}
\label{5}
\end{equation}
The higher the value of $H(p)$, the more uniform the probability distribution, which means higher uncertainty. However, sometimes the original entropy can not describe the actual uncertainty well. For two distributions $p(a)=[0.5,0.25,0.25]$ and $p(b)=[0.5,0.5,0]$, according to the formula we get $H(p(a))>H(p(b))$, but $p(b)$ is more uncertain than $p(a)$ for the classification scenario. 

A simple but effect adjustment is to use only the first top-\textit{k} probability values to calculate the entropy. Suppose $p_1, p_2,...,p_k$ are the first k values with the highest probability, then the top-\textit{k}-entropy is defined as

\begin{equation}
H_{top\mbox{-}k}(p)=-\sum_{i=1}^{k}\widetilde{p_i}log(\widetilde{p_i})\text{,}
\label{6}
\end{equation}
where $\widetilde{p_i}=p_i/\sum_{i=1}^{k}$. Then, the final confidence, i.e. sample weight $w$ is calculated by

\begin{equation}
w = 1-H_{top\mbox{-}k}(p)/log(k)\text{,}
\label{7}
\end{equation}
where the purpose of $H_{top\mbox{-}k}(p)/log(k)$ is to normalize the value to $0\sim1$. The larger the value of $w$, the higher the confidence is, which indicates that the prediction result of this sample is more reliable. Through the above process, the weights $w_{\Omega_{x}}$ for all training examples can be calculated for next stage.

The second stage executes the re-training by the weighted cross entropy loss which is expressed as

\begin{equation}
L_{CW}=-w\sum_{i=1}^{M}y_ilog(p_i).
\label{8}
\end{equation}
The training process is the same as the normal way and when the model converges, the optimal parameters are learned. The two-stage training readjusts the model attention to different samples, and the feature representations of high-confidence signals which provide more valid information are learned, so as to improve the robustness of the classifier.

\section{Simulation Results and Discussion}

\subsection{Dataset and Simulation Settings}
We adopt the RadioML2016.10a dataset as our main research object, which is generated by Tim et al. via GNU radio in \cite{b16}. The data is uniformly distributed across all 11 modulation types and the SNRs ranging from -20 dB to +18 dB with a step of 2 dB.  The samples are divided into training set and testing set by the ratio 8:2. Training and performance evaluation are performed on the Google Colaboratory equipped with one Tesla K80 GPU and 16 GB RAM. All of the networks utilize the Adam optimizer \cite{b17} with a learning rate of 0.001. The batch size of each iteration is 512 to make full use of computing resources. An early stop strategy with a patience value of 15 is also adopted to prevent potential overfitting.

\subsection{Experimental Results}

\begin{table}[tbp]
\caption{Accuracy Comparison of Different \textit{k} Values }
\begin{center}
\begin{tabular}{|c|c|c|c|c|}
\hline
                                                      & k=2     & k=3              & k=4     & w/o CW Loss \\ \hline
\begin{tabular}[c]{@{}c@{}}Average\\ Acc\end{tabular} & 62.12\% & \textbf{62.66\%} & 62.11\% & 62.09\%      \\ \hline
\begin{tabular}[c]{@{}c@{}}Max\\ Acc\end{tabular}     & 92.75\% & \textbf{92.79\%} & 92.37\% & 91.63\%      \\ \hline
\end{tabular}
\label{tab1}
\end{center}
\end{table}

First, a comparison experiment is conducted to determine the optimal value of hyper-parameter \textit{k} in the top-\textit{k} cross entropy. The average and maximum accuracy under all SNRs are listed in Table~\ref{tab1}. Based on the results, we adopt $k=3$ in the follow-up experiments. 
\begin{figure}[tbp]
\centerline{\includegraphics[scale=0.25]{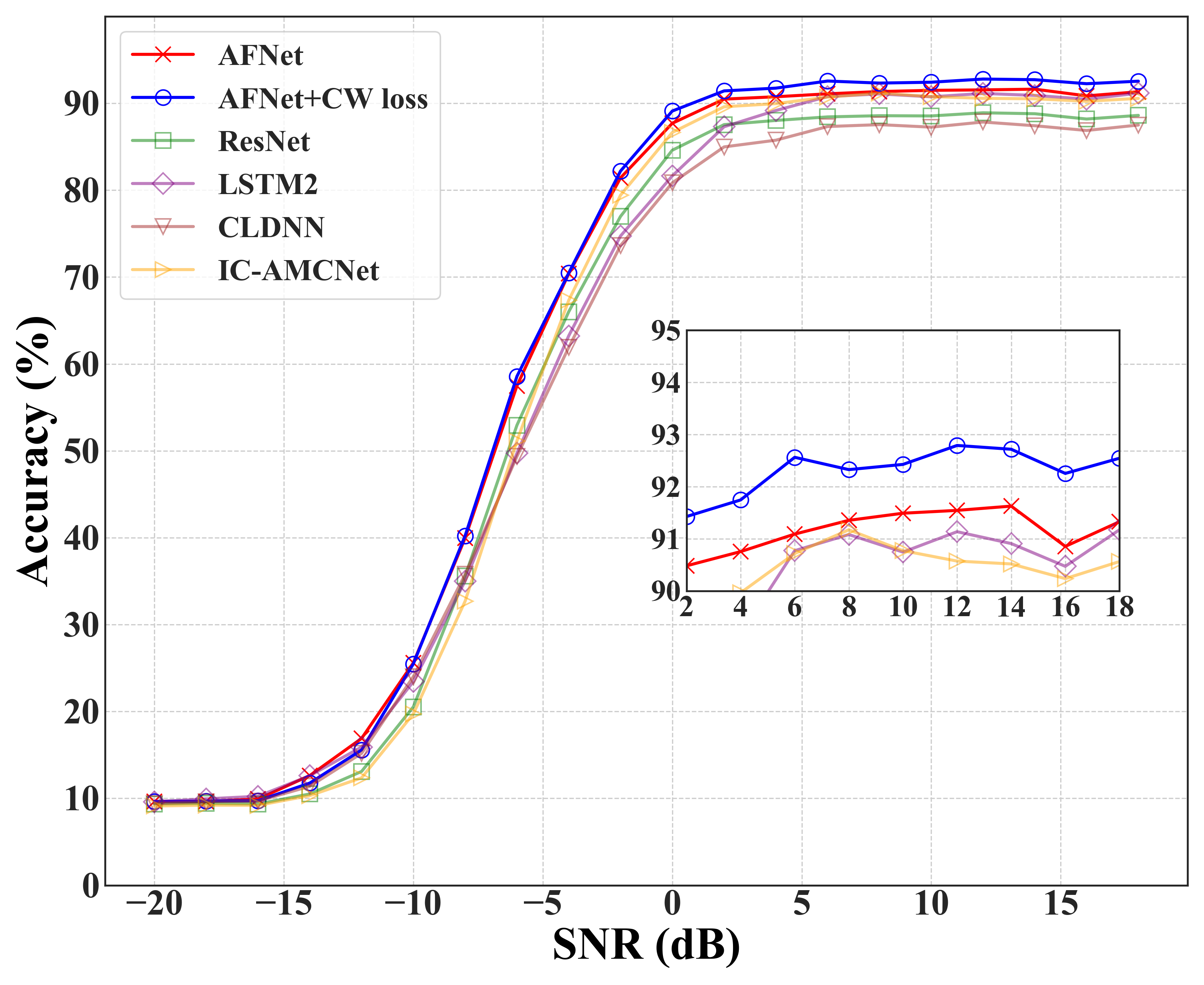}} 
\caption{Classification accuracy versus SNR of the proposed CW loss trained AFNet against other methods.} 
\label{fig4} 
\end{figure}

For exploring the overall performance of the proposed AMC scheme, comparative experiments are performed against other DL-based models, which are ResNet\cite{b6}, LSTM2\cite{b7}, CLDNN\cite{b9} and IC-AMCNet\cite{b10}. The classification accuracy on the same testing set of the proposed method and the other models at varying SNRs are detailed in Fig.~\ref{fig4}. It is observed that the proposed AFNet has outperformed other examined models. Moreover, the CW loss further improves its overall accuracy especially at high SNRs and the combined scheme achieves the superior performance. The results not only prove that the special design of AFNet is effective, but also indicate that the CW loss is conducive to improve the model's attention to high-quality samples and there is little accuracy degradation for low-quality samples.

\begin{figure}[tbp]
\centerline{\includegraphics[scale=0.18]{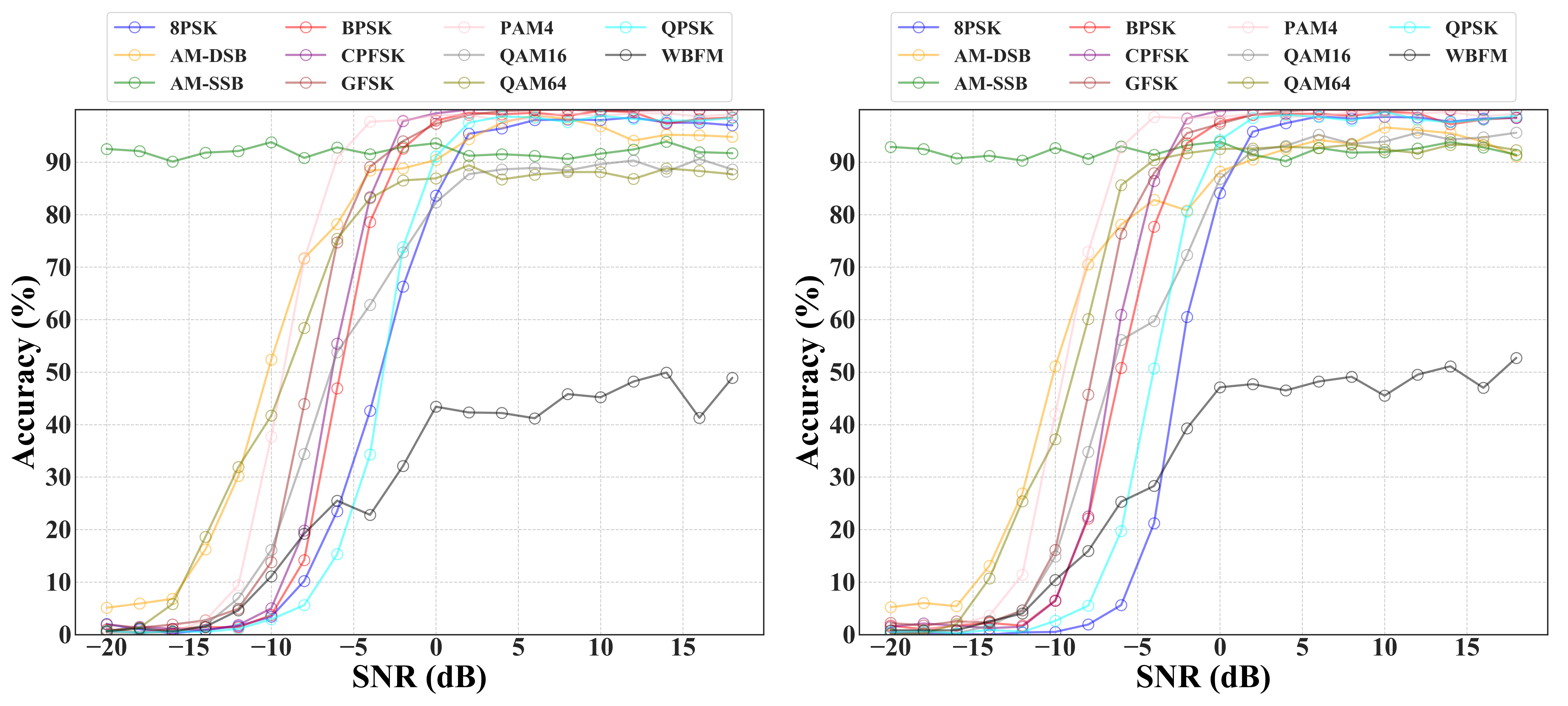}} 
\caption{Classification accuracy of different modulation schemes versus SNR. The left is the result of AFNet trained by vanilla CE loss and the right is obtained by the CW loss.} 
\label{fig5} 
\end{figure}

\begin{figure}[tbp]
\centerline{\includegraphics[scale=0.3]{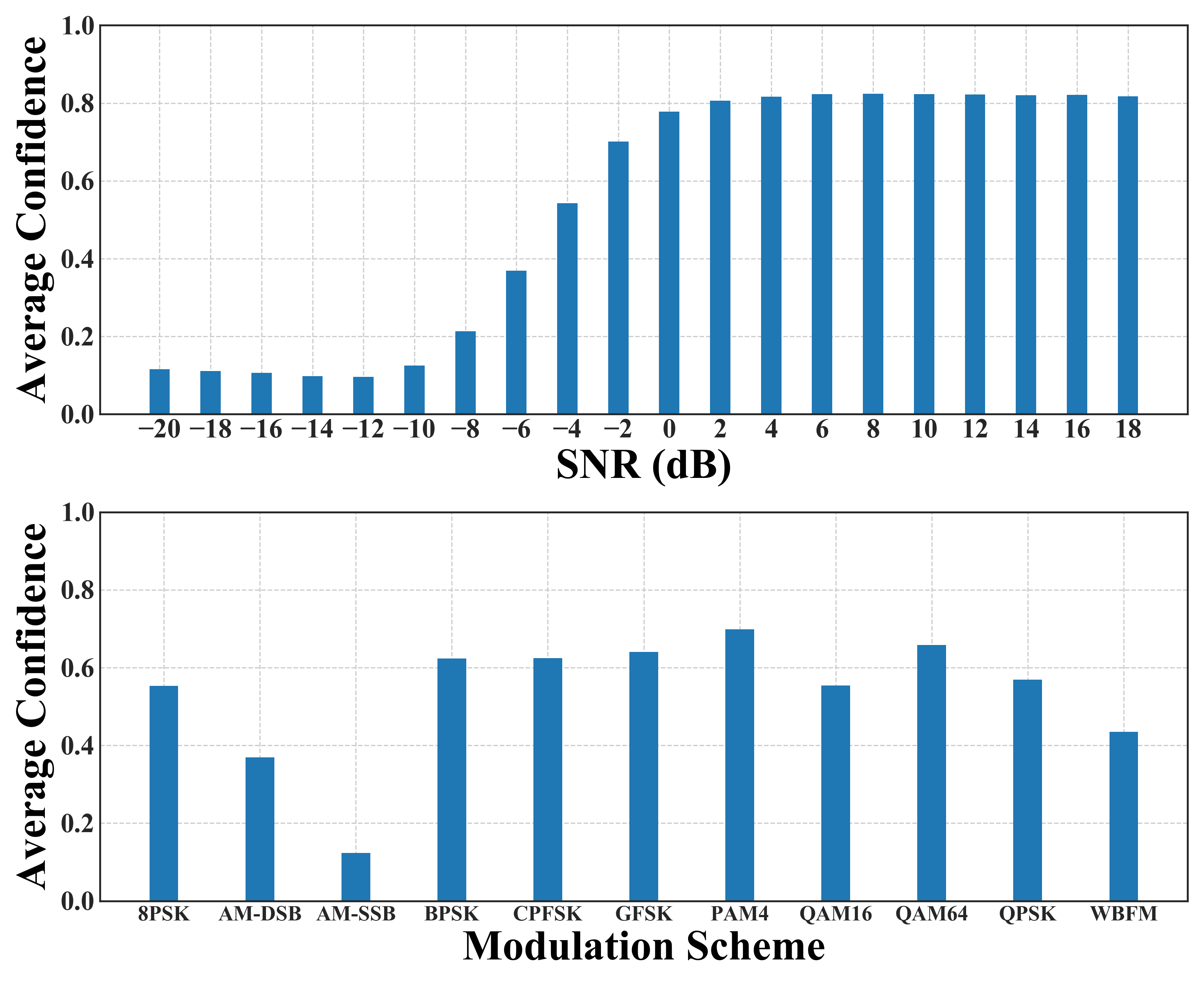}} 
\caption{Average instance confidence distribution for SNRs (upper) and modulation schemes (lower).} 
\label{fig6} 
\end{figure}

The classification accuracy curves of 11 modulation schemes are fully illustrated in Fig.~\ref{fig5}. Comparing the results before and after using CW loss, it can be seen that in general all high SNR signals are identified more accurately. From the perspective of modulation type, the recognition rates of QAM16, QAM64, QPSK and WBFM have been improved, along with certain decline of some other modulation types. Therefore, it can be inferred that the overall improvement of classification performance stems from the fact that the model has adjusted its prediction strategy to more rationally allocate its efforts to various signals.

In order to understand what kinds of samples are high-confidence and what are low-confidence in the cognition of deep models, we visualize the relationship between confidence and SNR and modulation type respectively, as presented in Fig.~\ref{fig6}. The results indicate that the average confidences of high SNR signals and digital modulation signals are relatively higher, which is consistent with the intuitive assumption.

Finally, in order to verify the generality of the proposed two-stage training and CW loss, the instance weights obtained by AFNet are utilized to re-train other models. The performance improvements are described in Table~\ref{tab2}.
The results show that CW loss brings different degrees of improvement to the classification performance of all models, among which ResNet achieves the largest gain. This fully illustrates the validity and potential of our proposed CW loss.

\begin{table}[tbp]
\caption{Performance of CW Loss on Other Models}
\begin{center}
\begin{tabular}{|c|c|c|}
\hline
                  & Average Acc                           & Max Acc        \\ \hline
ResNet            & 59.20\%                               & 88.90\%        \\
ResNet+CW loss    & {(\textbf{+1.95}) \textbf{61.15\%}}                    & (\textbf{+3.71}) \textbf{92.61\%} \\ \hline
CLDNN             & 58.17\%                               & 87.84\%        \\
CLDNN+CW loss     & (+1.02) 59.19\%                        & (+2.06) 89.90\% \\ \hline
LSTM2             & 59.94\%                               & 91.17\%        \\
LSTM2+CW loss     & (+0.25) 60.19\%                        & (+0.93) 92.10\% \\ \hline
IC-AMCNet         & 60.06\%                               & 90.77\%        \\
IC-AMCNet+CW loss & (+0.37) 60.43\%                        & (+0.88) 91.65\% \\ \hline
\end{tabular}
\label{tab2}
\end{center}
\end{table}

\section{Conclusion}

In this work, in order to improve the performance of AMC under noisy conditions, a novel DL-based scheme is proposed, which includes the  AFNet and CW loss function. The AFNet has a dynamic receptive field mechanism, which can adapt to different modulation schemes and capture the modulation information contained in adjacent symbols. The CW loss is exploited to improve the classification performance of high-confidence samples by re-weighting the samples according to the predicted class distribution. Simulation results show that the proposed scheme outperforms other DL-based models in terms of average and maximum accuracy. Moreover, the effectiveness of CW loss function with the two-stage learning strategy is investigated by implementing on different models and the significant improvements show its application potential.

\end{document}